\documentclass[conference,letterpaper]{IEEEtran}

\usepackage[utf8]{inputenc} 
\usepackage[T1]{fontenc}
\usepackage{url}
\usepackage{ifthen}
\usepackage{cite}
\usepackage[cmex10]{amsmath} 
\usepackage{graphicx}
\usepackage{subfig}

\usepackage{comment}
\usepackage{algorithm,algcompatible}
\algnewcommand\INPUT{\item[\textbf{Input:}]}%
\algnewcommand\OUTPUT{\item[\textbf{Output:}]}%
\usepackage[dvipsnames]{xcolor}
\usepackage{amsfonts}
\usepackage{amssymb}
\usepackage{comment}
\usepackage{latexsym}
\usepackage{bm}
\usepackage{bmpsize}
\usepackage[justification=centering]{caption}

\DeclareMathOperator*{\argmin}{arg\,min}
\DeclareGraphicsExtensions{.eps}
\graphicspath{{../jpg/}{../pdf/}{../jpeg/}{../eps/}}
\newtheorem{prop}{Proposition}

\newtheorem{lemma}{Lemma}

\begin{document}
\title{Using Quantization to Deploy Heterogeneous Nodes in Two-Tier Wireless Sensor Networks} 

\author{
  \IEEEauthorblockN{Saeed Karimi-Bidhendi, Jun Guo, Hamid Jafarkhani}
  \IEEEauthorblockA{Center for Pervasive Communications \& Computing\\
                    University of California, Irvine\\
                    Email: \{skarimib, guoj4, hamidj\}@uci.edu}
}

\maketitle

\begin{abstract}
We study a heterogeneous two-tier wireless sensor network in which $N$ heterogeneous access points (APs) collect sensing data from densely distributed sensors and then forward the data to $M$ heterogeneous fusion centers
(FCs). This heterogeneous node deployment problem is modeled as a quantization problem with distortion defined as the total power consumption of the network. The necessary conditions of the optimal AP and FC node deployment are explored in this paper. We provide a variation of Voronoi Diagram as the optimal cell partition for this network, and show that each AP should be placed between its connected FC and the geometric center of its cell partition. In addition, we propose a heterogeneous two-tier Lloyd
algorithm to optimize the node deployment. Simulation results show that our proposed algorithm outperforms the existing clustering methods like Minimum  Energy Routing, Agglomerative Clustering, and Divisive  Clustering, on average.
\end{abstract}

\begin{IEEEkeywords}
quantization, node deployment, heterogeneous wireless sensor networks.
\end{IEEEkeywords}

\section{Introduction}

Wireless sensor networks (WSNs) have been widely used to gather data from the environment and transfer the sensed information through wireless channels to one or more fusion centers. Based on the network architecture, WSNs can be classified as either non-hierarchical WSNs in which every sensor node has identical functionality and the connectivity of network is usually maintained by multi-hop wireless communications, or hierarchical WSNs where sensor nodes play different roles as they are often divided into clusters and some of them are selected as cluster head or relay. WSNs can also be divided into either homogeneous WSNs \cite{JG,SD,AS,MBMOS}, in which sensors share the same capacity, e.g., storage, computation power, antennas, sensitivity etc., or heterogeneous WSNs where sensors have different capacities \cite{YTH,JG2,JPH,MNMA}.

Energy consumption is a key bottleneck in WSNs due to limited energy resources of sensors, and difficulty or even infeasibility of recharging the batteries of densely deployed sensors. The energy consumption of a sensor node comes from three primary components: communication energy, computation energy and sensing energy. The experimental measurements show that, in many applications, the computation energy is negligible when it is compared to communication energy\cite{GMMA, MS}. Furthermore, for passive sensors, such as light sensors and acceleration sensors, the sensing energy is significantly small. Therefore, wireless communication dominates the sensor energy consumption in practice. There are three primary methods to reduce the energy consumption of radio communication in the literature: (1) topology control\cite{XYY,XW}, in which unnecessary energy consumption is avoided by properly switching awake and asleep states,  (2) energy-efficient routing protocols \cite{MBMOS,JGBC}, that are designed to find an optimal path to transfer data, (3) power control protocols \cite{VP,SVRP}, that save communication energy by adjusting the transmitter power at each node while keeping reliable communications. Another widely used method, Clustering \cite{OYSF,VP}, attempts to balance the energy consumption among sensor nodes by iteratively selecting cluster heads. Unfortunately, above MAC protocols bring about a massive number of message exchanges because the geometry and/or energy information are required during the operation \cite{OYSF}. Also, the node deployment is known and fixed in the aforementioned energy saving approaches while it plays an important role in energy consumption of the WSNs. 

In this paper, we study the node deployment problem in heterogeneous two-tier WSNs consisting of heterogeneous APs and heterogeneous FCs, with distortion defined as the total wireless communication power consumption. The optimal energy-efficient sensor deployment in homogeneous WSNs is studied in \cite{JG}. However, the homogeneous two-tier WSNs in \cite{JG} do not address various challenges that exist in the heterogeneous two-tier WSNs, e.g., unlike regular Voronoi diagrams for homogeneous WSNs, the optimal cells in heterogeneous WSNs may be non-convex, not star-shaped or even disconnected, and the cell boundaries may not be hyperplanes. Another challenge that exists in the heterogeneous two-tier networks is that unlike the homogeneous case \cite{JG}, or heterogeneous one-tier case \cite{On the minimum average distortion of quantizers with index-dependent}, some nodes may not contribute to the energy saving. To the best of our knowledge, the optimal node deployment for energy efficiency in heterogeneous WSNs is still an open problem. Our main goal is to find the optimal AP and FC deployment to minimize the total communication energy consumption. By deriving the necessary conditions of the optimal deployments in such heterogeneous two-tier WSNs, we design a Lloyd-like algorithm to deploy nodes.

The rest of this paper is organized as follows: In Section \ref{sec:model}, we introduce the system model and problem formulation. In Section \ref{sec:opt}, we study the optimal AP and FC deployment. A numerical algorithm is proposed in Section \ref{sec:algorithm} to minimize the energy consumption. Section \ref{sec:simulation} presents the experimental results and Section \ref{sec:conclusion} concludes the paper.

\section{System Model and Problem Formulation}\label{sec:model}

Here, we study the power consumption of the heterogeneous two-tier WSNs consisting of three types of nodes, i.e., homogeneous sensors, heterogeneous APs and heterogeneous FCs. The power consumption models for homogeneous WSNs are discussed in details in \cite{JG}. The main difference in this work is the heterogeneous characteristics of the APs and FCs. For the sake of completeness, we describe the system model for heterogeneous WSNs here in details. Given the target area $\Omega\subset \mathbb{R}^2$ which is a convex polygon including its interior, $N$ APs and $M$ FCs are deployed to gather data from sensors. Throughout this paper, we assume that $N\geq M$. Given the sets of AP and FC indices, i.e., $\mathcal{I_A}=\{1,2,..., N\}$ and $\mathcal{I_B}=\{1,2,..., M\}$, respectively, the index map $T:\mathcal{I_A} \longrightarrow \mathcal{I_B}$ is defined to be $T(n)=m$ if and only if AP $n$ is connected to FC $m$. The AP and FC deployments are then defined by $P=\left(p_1, ..., p_N \right)$ and $Q=\left(q_1, ..., q_M \right)$, where $p_n, q_m\in \mathbb{R}^2$ denote the location of AP $n$ and FC $m$, respectively. Throughout this paper, we assume that each sensor only sends data to one AP. For each $n\in \mathcal{I_A}$, AP $n$ collects data from sensors in the region $\emph{R}_n \subset \Omega$; therefore, for each AP deployment $P$, there exists an AP partition $\mathbf{R}=(\emph{R}_1,...,\emph{R}_N)$ comprised of disjoint subsets of $\mathbb{R}^2$ whose union is $\Omega$. The density of the data rate from the densely distributed sensors is denoted via a continuous and differentiable function $f: \Omega \longrightarrow \mathbb{R}^+$, i.e., the total amount of data gathered from the sensors in region $\emph{R}_n$ in one time unit is $\int_{\emph{R}_n}f(w)dw$ \cite{JG}.

We focus on the power consumption of sensors and APs, since FCs usually have reliable energy resources and their energy consumption is not the main concern. First, we discuss the APs' total power consumption. According to \cite{YTH}, power at the receiver of AP $n$ is modeled as $\mathcal{P}_{r_n}^{A}=\rho_n\int_{R_n}f(w)dw, n\in \mathcal{I}_{A}$, where $\rho_n$ is AP $n$'s power consumption coefficient for receiving data. For simplicity, we assume APs share the same receiving coefficient, i.e., $\rho_n=\rho$. Therefore, the sum of power consumption at receivers is a constant and does not affect the energy optimization and can be ignored in our objective function. In what follows, we focus on power consumption at AP transmitters. The \emph{average-transmitting-power} (Watts) of AP $n$ is defined to be $\mathcal{P}_{t_n}^{\mathcal{A}} = E_{t_n}^{\mathcal{A}}\Gamma_n^{\mathcal{A}}, \forall n\in \mathcal{I_A}$, where $E_{t_n}^{\mathcal{A}}$ denotes the \emph{instant-transmission-power} (Joules/second) of AP $n$, and $\Gamma_n^{\mathcal{A}}$ denotes the \emph{channel-busy-ratio} for the channel of AP $n$ to its corresponding FC, i.e., the percentage of time that the transmitter forwards data. According to \cite[(2.7)]{AG}, the instant-receive-power through free space is  $\mathcal{P}_{r_n}^{\mathcal{A}}=\frac{\mathcal{P}_{r_n}^{\mathcal{A}}G_tG_r\lambda^2}{16\pi^2d^2}$, where $G_t$ is the transmitter antenna gain, $G_r$ is the receiver antenna gain, $\lambda$ is the signal wavelength, and $d$ is the distance between the transmit and receive antennas. Let $\mathcal{N}_0$ be the noise power. In order to achieve the required SNR threshold $\gamma$ at the receivers, i.e., $\frac{\mathcal{P}_{r_n}^{\mathcal{A}}}{\mathcal{N}_0} = \gamma$, the \emph{instant-transmission-power} from AP $n$ to FC $T(n)$ should be set to $E_{t_n}^{\mathcal{A}} = \eta_{n, T(n)}^{\mathcal{A}}||p_n-q_{T(n)}||^2$, where $||.||$ denotes the Euclidean distance, $\eta_{n, T(n)}^{\mathcal{A}}$ is a constant determined by the antenna gain of AP $n$ and the SNR threshold of FC $T(n)$. Since AP $n$ gathers data from the sensors in $R_n$, the amount of data received from sensors in one time unit, i.e. the \emph{average-receiver-data-rate}, is $\int_{R_n}f(w)dw$. It can be reasonably assumed that the AP transmitters only forward sensing data when the collected data comes into the buffer. Therefore, the \emph{channel-busy-ratio} is proportional to the \emph{average-receiver-data-rate}, and can be written as $\Gamma_n^{\mathcal{A}}=\frac{\int_{R_n}f(w)dw.T/\zeta^{\mathcal{A}}_{T(n)}}{T} = \frac{\int_{R_n}f(w)dw}{\zeta^{\mathcal{A}}_{T(n)}}$, where $\zeta_{ T(n)}^{\mathcal{A}}$ is the AP $n$'s \emph{instant-transmitter-data-rate} which is determined by the SNR threshold at the corresponding FC $T(n)$. Hence, we can rewrite the \emph{average-transmitter-power} of AP $n$ as $\mathcal{P}_{t_n}^{\mathcal{A}}=E_{t_n}^\mathcal{A}\Gamma_n^\mathcal{A}=\frac{\eta_{n,T(n)}^{\mathcal{A}}}{\zeta_{T(n)}^{\mathcal{A}}}||p_n-q_{T(n)}||^2\int_{R_n}f(w)dw$, and the total power consumption at AP transmitters is calculated by summing the \emph{average-transmitter-powers} of APs:
\begin{align}\label{eq1}
         \overline{\mathcal{P}}^{\mathcal{A}}\left(P, Q, \mathbf{R}, T \right) &=  \sum_{n=1}^{N}\mathcal{P}_{t_n}^{\mathcal{A}} \\ &=  \sum_{n=1}^{N}\int_{R_n}\frac{\eta_{n,T(n)}^{\mathcal{A}}}{\zeta_{T(n)}^{\mathcal{A}}}||p_n-q_{T(n)}||^2f(w)dw    \nonumber
\end{align}

Second, we consider sensors' total transmitting power consumption. The total amount of data collected from the sensors inside the region $[w,w+dw]$ in one time unit is equal to $f(w)dw$ since the density of data rate $f(.)$ is approximately uniform on the extremely small region $[w,w+dw]$. Therefore, the sum of \emph{channel-busy-ratios} of sensors in the infinitesimal region $[w,w+dw]$ is $\Gamma_n^{\mathcal{S}} = \frac{f(w)dw.T/\zeta_n^{\mathcal{S}}}{T} = \frac{f(w)dw}{\zeta_n^{\mathcal{S}}}$, where $\zeta_n^{\mathcal{S}}$ is sensors \emph{instant-transmitter-data-rate}. We only consider the homogeneous sensors, i.e., sensors' antenna gains are identical. Moreover, sensors within $[w,w+dw]$ have approximately the same distance to the corresponding AP $p_n$, and thus have the same \emph{instant-transmission-power} $E_{t_n}^{\mathcal{S}}=\eta_n^{\mathcal{S}}||w-p_n||^2$, where $\eta_n^{\mathcal{S}}$ is a constant determined by sensors' common transmitter antenna gain, AP $n$'s receiver antenna gain, and the SNR threshold of AP $n$. Therefore, the sum of \emph{average-transmitter-powers} within the region $[w,w+dw]$ is equal to $\frac{\eta_n^\mathcal{S}}{\zeta_n^\mathcal{S}}||p_n - w||^2f(w)dw$. Since sensors in the region $R_n$ send their data to AP $n$, the sum \emph{average-transmitter-powers} of sensors in the target area $\Omega$ can be written as:
\begin{equation}\label{eq2}
    \overline{\mathcal{P}}^{\mathcal{S}}(P, \mathbf{R}) = \sum_{n=1}^N\int_{R_n}\frac{\eta_n^\mathcal{S}}{\zeta_n^\mathcal{S}}||p_n - w||^2f(w)dw
\end{equation}

The two-tier distortion is then defined as the Lagrangian function of Eqs. (\ref{eq1}) and (\ref{eq2}):
\begin{align}\label{eq4}
        &D\left(P,Q, \mathbf{R}, T \right) = \overline{\mathcal{P}}^\mathcal{S}\left(P, \mathbf{R} \right) + \beta \overline{\mathcal{P}}^\mathcal{A}\left(P,Q, \mathbf{R}, T \right) = \\
&\sum_{n=1}^N\int_{R_n}\left(a_n||p_n-w||^2 + \beta b_{n,T(n)}||p_n-q_{T(n)}||^2 \right)f(w)dw \nonumber
\end{align}
where $a_n=\frac{\eta_n^\mathcal{S}}{\zeta_n^{\mathcal{S}}}$ and $b_{n,T(n)}=\frac{\eta_{n,T(n)}^\mathcal{A}}{\zeta_{T(n)}^\mathcal{A}}$. Our main objective in this paper is to minimize the two-tier distortion defined in Eq. (\ref{eq4}) over the AP deployment $P$, FC deployment $Q$, cell partition $\mathbf{R}$ and index map $T$.

\section{Optimal Node Deployment in Two-Tier WSNs}\label{sec:opt}

Let the optimal AP and FC deployments, cell partitions and index map be denoted by $P^*=\left(p_1^*,...,p_N^* \right)$, $Q^*=\left(q_1^*,...,q_M^* \right)$, $\mathbf{R}^*=(R^*_1,...,R^*_N)$ and $T^*$, respectively. In what follows, we determine the properties of such an optimal node deployment $\left(P^*,Q^*,\mathbf{R}^*,T^* \right)$ that minimizes the two-tier distortion in (\ref{eq4}). Note that the index map only appears in the second term of Eq. (\ref{eq4}); thus, for any given AP and FC deployment $P$ and $Q$, the optimal index map is given by:
\begin{equation}\label{eq8}
    T_{[P,Q]}(n) = \argmin_m b_{n,m}||p_n - q_m||^2
\end{equation}
Eq. (\ref{eq8}) implies that an AP may not be connected to its closest FC due to heterogeneity of the APs and FCs, and to minimize the two-tier distortion, AP $n$ should be connected to FC $m$ that minimizes the weighted distance $b_{n,m}||p_n - q_m||^2$. Inspired by definition of the two-tier distortion in (\ref{eq4}), for each $n\in \mathcal{I_A}$, the Voronoi cell $V_n$ for AP and FC deployments $P$ and $Q$, and index map $T$ is defined as:
\begin{multline}\label{eq5}
    V_n(P,Q,T) \triangleq \{w: a_n||p_n-w||^2 + \beta b_{n, T(n)}||p_n-q_{T(n)}||^2 \\ \leq  a_k||p_k-w||^2 + \beta b_{k, T(k)}||p_k-q_{T(k)}||^2, \forall k \neq n \}
\end{multline}
Ties are broken in favor of the smaller index to ensure that each Voronoi cell $V_n$ is a Borel set. When it is clear from context, we write $V_n$ instead of $V_n(P,Q,T)$. The collection
\begin{equation}\label{eq6}
\mathbf{V}(P,Q,T) \triangleq (V_1,V_2,...,V_N)
\end{equation}
is referred to as the generalized Voronoi diagram. Note that unlike the regular Voronoi diagrams, the Voronoi cells defined in Eq. (\ref{eq5}) may be non-convex, not star-shaped or even disconnected. The following proposition establishes that the generalized Voronoi diagram in (\ref{eq6}) provides the optimal cell partitions, i.e., $\mathbf{R}^*(P,Q,T) = \mathbf{V}(P,Q,T)$.
\begin{prop}\label{allactive}
For any partition of the target area $\Omega$ such as $U$, and any AP and FC node deployments such as $P$ and $Q$ and each index map $T$ we have:
\begin{equation}\label{eq7}
    D\left(P,Q,U,T \right) \geq D\left(P,Q,\mathbf{V}(P,Q,T), T \right)
\end{equation}
\end{prop}
The proof is provided in Appendix A. Note that given AP and FC deployments $P$ and $Q$, the optimal index map and cell partitioning can then be determined by Eqs. (\ref{eq8}) and (\ref{eq6}). The following lemma demonstrates that in any optimal node deployment $\left(P^*, Q^*, \mathbf{R}^*, T^* \right)$, each FC contributes to the total distortion, i.e., adding an additional FC results in a strictly lower optimal two-tier distortion regardless of its weights $b_{n,M+1}$ as long as $M<N$ holds.
\begin{lemma}\label{lemma1}
Let $\left(P^*, Q^*, \mathbf{R}^*, T^* \right)$ be the optimal node deployment for $N$ APs and $M$ FCs. Given an additional FC with parameters $b_{n,M+1}$ for every $n\in \mathcal{I_A}$, the optimal AP and FC deployments, index map and cell partitioning are denoted via $P'=\left(p_1',p_2',...,p_N'\right)$, $Q'=\left(q_1',q_2',...,q_{M+1}'\right)$, $T'$ and $\mathbf{R}'$, respectively. Assuming $M<N$, we have:
\begin{equation}\label{FC_Decrease_Distortion}
D\left(P',Q',\mathbf{R}',T' \right) < D\left(P^*,Q^*,\mathbf{R}^*,T^* \right)
\end{equation}
\end{lemma}
The proof is provided in Appendix B. While Lemma \ref{lemma1} indicates that each FC contributes to the distortion, same may not hold for some APs. As an example to show the existence of useless APs in the optimal deployment, consider two APs and one FC and one-dimensional target region $\Omega=[0,1]$ with parameters $a_1=b_1=1$, $a_2=b_2=100$, $\beta=1$ and a uniform density function. We search the optimal deployments by Brute-force search. According to our simulation, the optimal deployments share the following properties: (i) Both FC and AP 1 are placed at the centroid of the target region, i.e., $q^*_1=p^*_1=0.5$; (ii) AP 2's partition is empty, i.e., $V_2(P^*,Q^*,T_{[P^*,Q^*]})=\varnothing$. Property (ii) implies that AP $2$ does not contribute to the two-tier distortion of optimal node deployment. Let $v_n^*(P,Q,T)=\int_{R_n^*}f(w)dw$ be the Lebesgue measure (volume) of the region $R_n^*$, and $c_n^*(P,Q,T) = \frac{\int_{R_n^*}wf(w)dw}{\int_{R_n^*}f(w)dw}$ be the geometric centroid of the region $R_n^*$. When there is no ambiguity, we write $v_n^*(P,Q,T)$ and $c_n^*(P,Q,T)$ as $v_n^*$ and $c_n^*$, respectively. Lemma \ref{lemma1} immediately leads to the following corollary.

\emph{Corollary 1. }Let $\left(P^*, Q^*, \mathbf{R}^*, T^* \right)$ be the optimal node deployment for $N$ APs and $M$ FCs. If $M\leq N$, then for each $m\in \mathcal{I_B}$, $\sum_{n:T^*(n)=m}v^*_n>0$.

The proof can be found in Appendix C. The following proposition provides the necessary conditions for the optimal AP and FC deployments in the heterogeneous two-tier WSNs.
\begin{prop}\label{necessary}
The necessary conditions for optimal deployments in the heterogeneous two-tier WSNs with the distortion defined in (\ref{eq4}) are:
\begin{equation}\label{eq9}
    \begin{split}
        p_n^* &= \frac{a_n c_n^* + \beta b_{n,T^*(n)}q^*_{T^*(n)}}{a_n + \beta b_{n,T^*(n)}} \\
        q^*_m & = \frac{\sum_{n:T^*(n)=m}{}b_{n,m}p^*_nv^*_n}{\sum_{n:T^*(n)=m}{}b_{n,m}v^*_n}
    \end{split}
\end{equation}
\end{prop}
The proof is provided in Appendix D. Corollary 1 implies that the denumerator of the second equation in (\ref{eq9}) is positive; thus, $q^*_m$ is well-defined. According to Eq. (\ref{eq9}), the optimal location of FC $m$ is the linear combination of the locations of its connected APs, and the optimal location of AP $n$ is on the segment $\overline{c^*_nq^*_{T^*(n)}}$. In the next section, we use the properties derived in Propositions 1 and 2 and in Eq. (\ref{eq8}), and design a Lloyd-like algorithm to find the optimal node deployment.

\section{Node Deployment Algorithm}\label{sec:algorithm}

First, we quickly review the conventional Lloyd algorithm. Lloyd algorithm iterates between two steps: In the first step, the node deployment is optimized while the partitioning is fixed and in the second step, the partitioning is optimized while the node deployment is fixed. Although the conventional Lloyd Algorithm can be used to solve one-tier quantizers or one-tier node deployment problems as shown in \cite{JG2}, it cannot be applied to two-tier WSNs where two kinds of nodes are deployed. Inspired by the properties explored in Section III, we propose a heterogeneous two-tier Lloyd (HTTL) algorithm to solve the optimal deployment problem in heterogeneous two-tier WSNs and minimize the two-tier distortion defined in (\ref{eq4}). Starting with a random initialization for node deployment $\left(P,Q,\mathbf{R},T \right)$ in the target area $\Omega$, our algorithm iterates between four steps: (i) Update the index map $T$ according to Eq. (\ref{eq8}); (ii) Obtain the cell partitioning according to Eq. (\ref{eq5}) and update the value of volumes $v_n$ and centroids $c_n$; (iii) Update the location of FCs according to Eq. (\ref{eq9}); (iv) Update the location of APs according to Eq. (\ref{eq9}). The algorithm continues until convergence. In Appendix E, we prove that the two-tier distortion will converge with HTTL algorithm. This procedure is summarized in Algorithm 1 below.

\begin{algorithm}
    \caption{HTTL Algorithm}
  \begin{algorithmic}[1]
    \INPUT Weights $\left\{a_n \right\}_{n\in \mathcal{I_A}}$ and $\left\{b_{n,m} \right\}_{n\in \mathcal{I_A},m\in \mathcal{I_B}}$. $\epsilon \in \mathbb{R}^+$.
    \OUTPUT Optimal node deployment $\left(P^*, Q^*, \mathbf{R}^*, T^* \right)$.
    \STATE Randomly initialize the node deployment $\left(P, Q, \mathbf{R}, T \right)$.
    
    \STATE {\bf do}
    \STATE Compute the two-tier distortion $D_{\textrm{old}}=D(P,Q,\mathbf{R},T)$.
    \STATE Update the index map $T$ according to Eq. (\ref{eq8}).
    \STATE Update the AP partitioning $\mathbf{R}$ by selecting its $n$th region as the generalized Voronoi region in (\ref{eq5}).
    \STATE Calculate the volumes $\{v_n\}_{n\in \mathcal{I_A}}$ and centroids $\{c_n\}_{n\in \mathcal{I_A}}$ of the AP partitioning $\mathbf{R}$.
    \STATE For each $m\in \mathcal{I_B}$, move the FC $m$ to $\frac{\sum_{n:T(n)=m}{}b_{n,m}p_nv_n}{\sum_{n:T(n)=m}{}b_{n,m}v_n}$.
    \STATE For each $n\in \mathcal{I_A}$, move the AP $n$ to $\frac{a_n c_n + \beta b_{n,T(n)}q_{T(n)}}{a_n + \beta b_{n,T(n)}}$.
    \STATE Update the two-tier distortion $D_{\textrm{new}}=D(P,Q,\mathbf{R},T)$.
    \STATE {\bf While }$\frac{D_{\textrm{old}} - D_{\textrm{new}}}{D_{\textrm{old}}}\geq \epsilon$
    \STATE {\bf Return:} The node deployment $\left(P,Q,\mathbf{R},T \right)$.
  \end{algorithmic}
\end{algorithm}

\section{Experiments}\label{sec:simulation}

We provide the experimental results in two heterogeneous two-tier WSNs: (i) WSN1: A heterogeneous WSN including 1 FC and 20 APs; (ii) WSN2: A heterogeneous WSN including 4 FCs and 20 APs. We consider the same target domain $\Omega$ as in \cite{JG, JPH}, i.e., $\Omega=[0,10]^2$. The data rate density function is set to a uniform function, $f(\omega) = \frac{1}{\int_{\Omega}dA} = 0.01$. To evaluate the performance, 10 initial AP and FC deployments on $\Omega$ are generated randomly, i.e, every node location is generated with uniform distribution on $\Omega$. In order to make a fair comparison to prior works, similar to the experimental setting in \cite{JG, JPH}, the maximum number of iterations is set to 100, FCs, APs, and geometric centroid of AP cells are denoted, respectively, by colored five-pointed stars, colored circles, and colored crosses. Other parameters are provided in Table \ref{SPT}. According to the parameters in Table \ref{SPT}, we divide APs into two groups: strong APs ($n\in\{1,\dots,10\}$) and weak APs ($n\in\{11,\dots,20\}$). Similarly, FCs are divided in strong FCs ($m\in\{1,2\}$) and weak FCs ($m\in\{3,4\}$). To distinguish strong APs (or FCs) and weak APs (or FCs), we denote strong and weak nodes by solid and hollow symbols, respectively.

\vspace{6pt}
\setlength{\intextsep}{0pt plus 0pt minus 6pt}
\setlength{\textfloatsep} {0pt plus 2pt minus 6pt}
\begin{table}[!bth]
\setlength\abovecaptionskip{0pt}
\setlength\belowcaptionskip{0pt}
\centering
\caption{Simulation Parameters}
\begin{tabular}{|c|c|c|c|c|c|c|c|c|c|}
\hline
Parameters & \!\!$a_{1:10}$\!\! & \!\!$a_{11:20}$\!\! & \!\!$b_{1:4,1:2}$\!\! & \!\!$b_{1:4,3:4}$\!\! & \!\!$b_{5:20,1:2}$\!\! & \!\!$b_{5:20,3:4}$\!\!\\
\hline
Values & 1 & 2 & 1 & 2 & 2 & 4\\
\hline
\end{tabular}
\label{SPT}
\end{table}
\vspace{5pt}

Like the experiments in \cite{JG}, we compare the weighted power of our proposed algorithm with Minimum Energy Routing (MER) \cite{AG}, Agglomerative Clustering (AC) \cite{DM}, and Divisive Clustering (DC) \cite{DM} algorithms. AC and DC are bottom-up and top-down clustering algorithms, respectively. MER is a combination of Multiplicatively weighted Voronoi Partition \cite{VD} and Bellman-Ford algorithms \cite[Section 2.3.4]{BG}. More details about MER, AC, and DC can be found in \cite{JG}.

\begin{figure}[!htb]
\setlength\abovecaptionskip{0pt}
\setlength\belowcaptionskip{0pt}
\centering
\subfloat[]{\includegraphics[width=43mm]{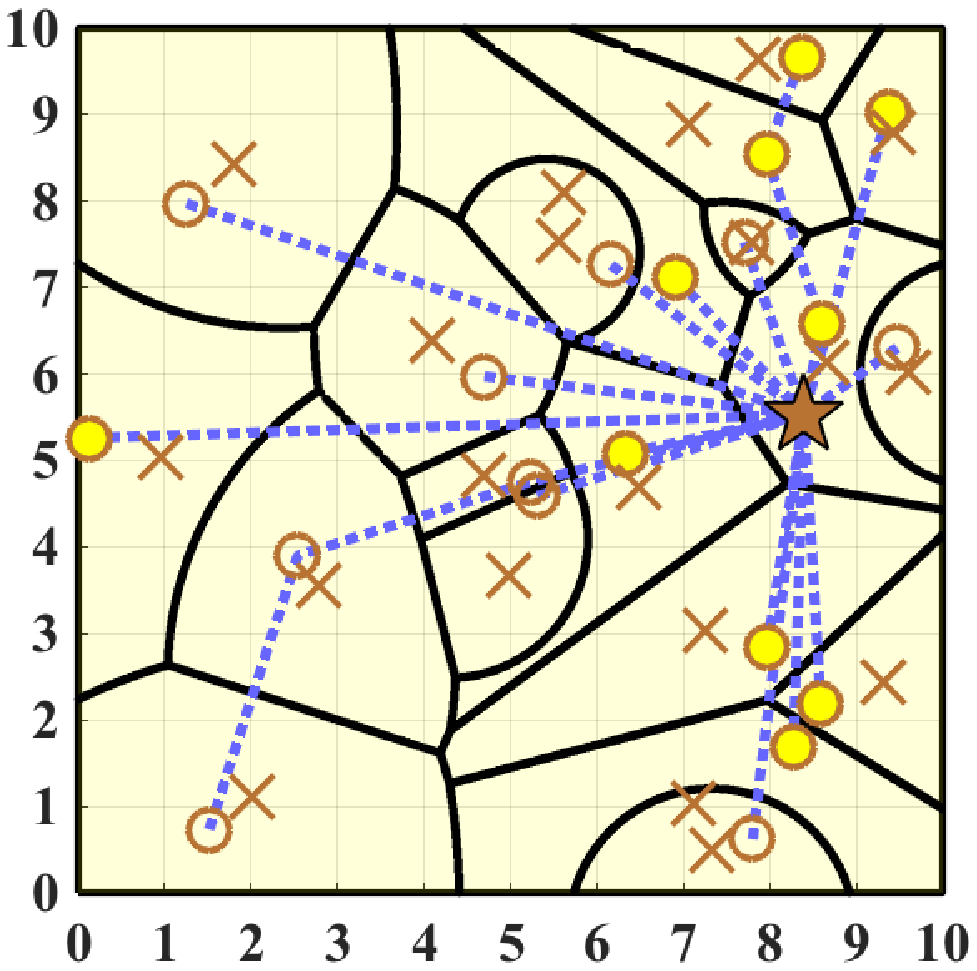}
\label{MERDeploymentInWSN1}}
\hfil
\subfloat[]{\includegraphics[width=43mm]{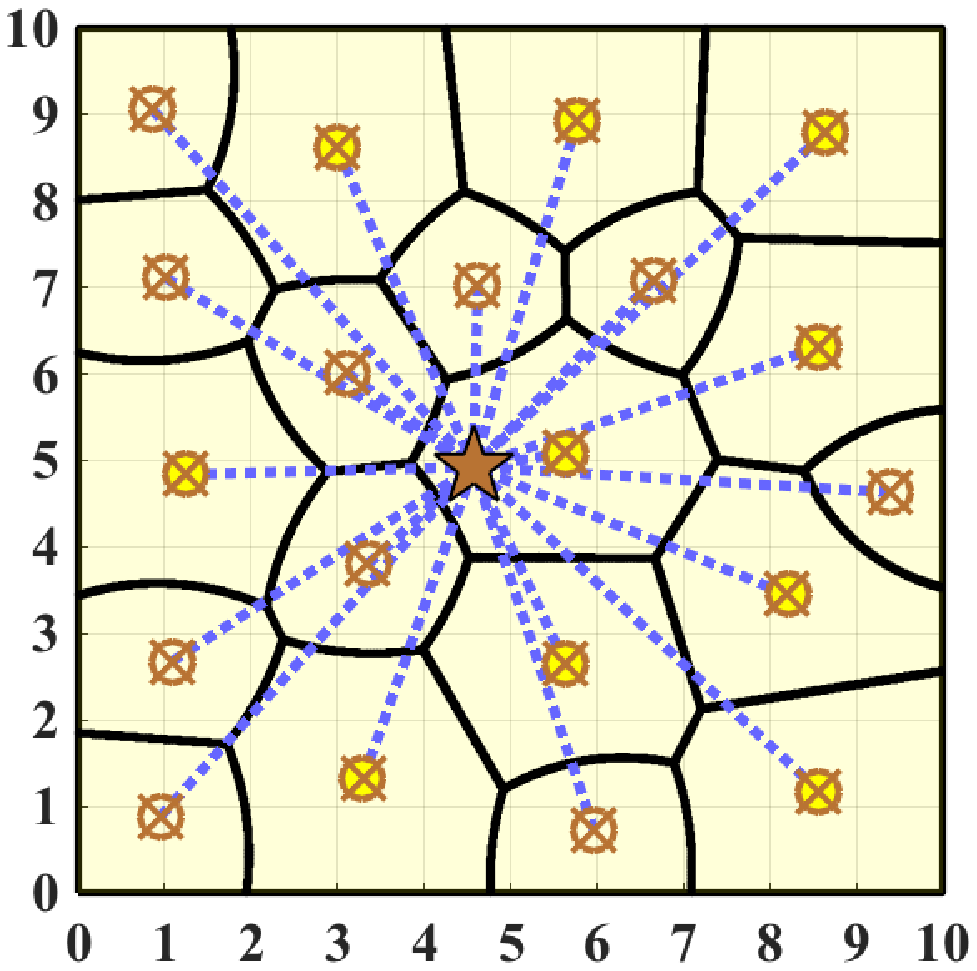}
\label{ACDeploymentInWSN1}}
\hfil
\subfloat[]{\includegraphics[width=43mm]{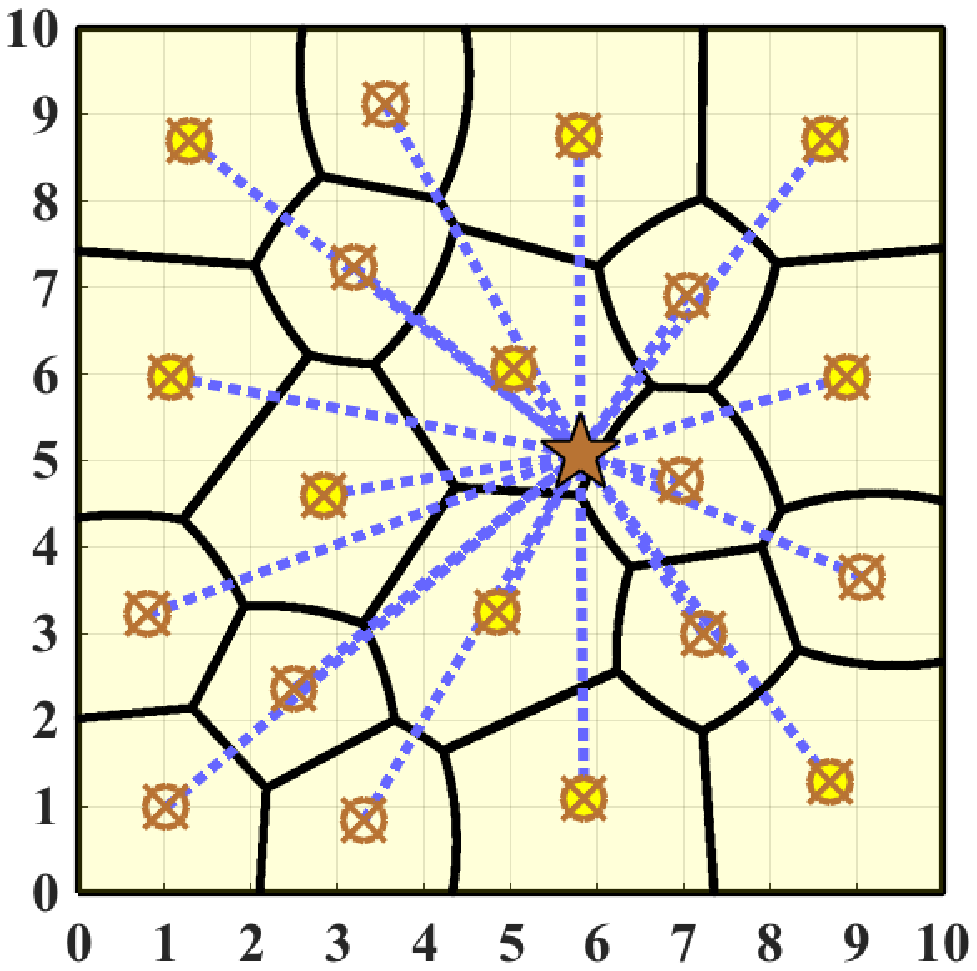}
\label{DCDeploymentInWSN1}}
\hfil
\subfloat[]{\includegraphics[width=43mm]{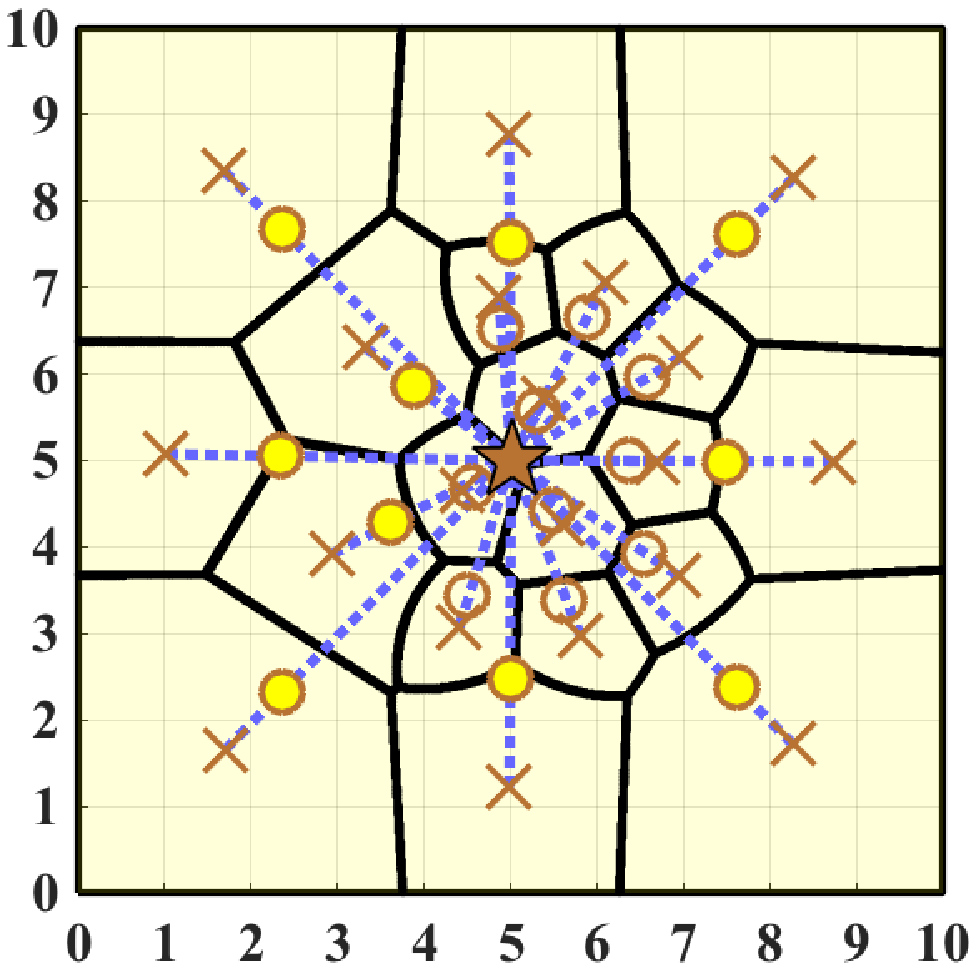}
\label{HELDeploymentInWSN1}}
\captionsetup{justification=justified}
\caption{\small{AP and FC deployments of different algorithms with $\beta=0.25$ in WSN1. (a) MER. (b) AC (c) DC. (d) HTTL.}}
\label{DeploymentInWSN1}
\end{figure}

\begin{figure}[!htb]
\setlength\abovecaptionskip{0pt}
\setlength\belowcaptionskip{0pt}
\centering
\subfloat[]{\includegraphics[width=43mm]{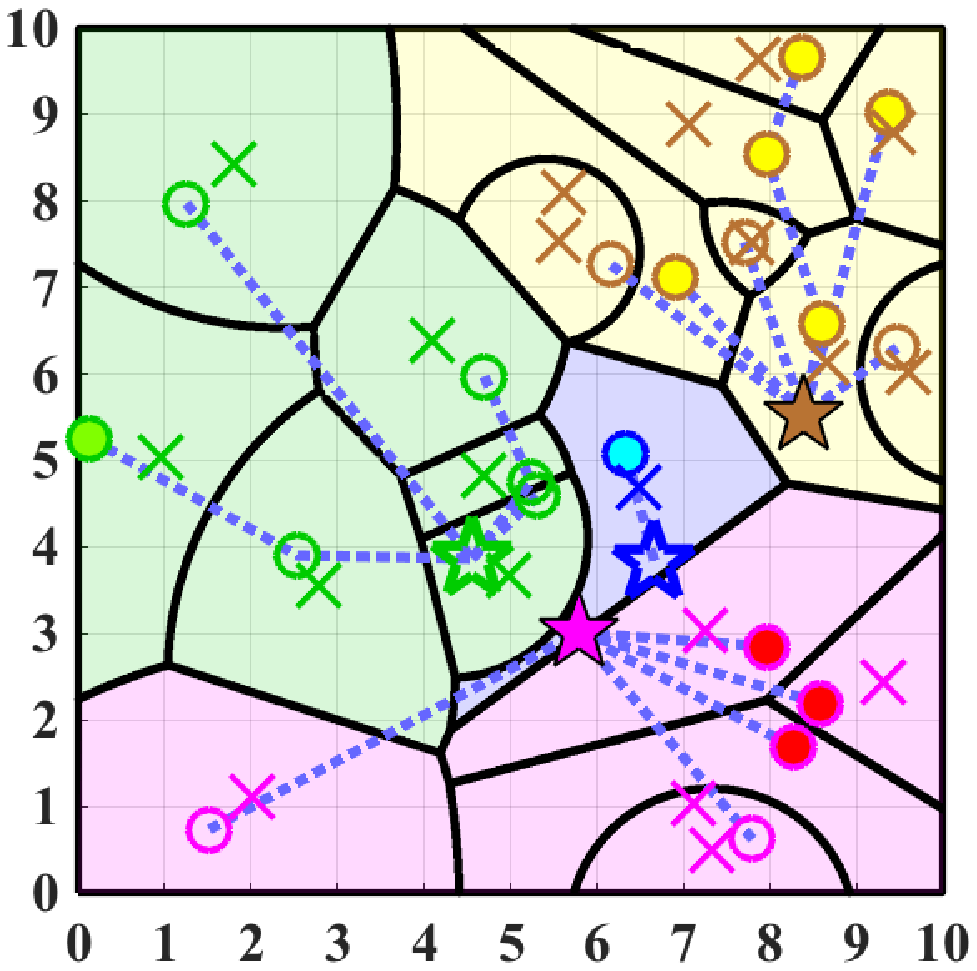}
\label{MERDeploymentInWSN2}}
\hfil
\subfloat[]{\includegraphics[width=43mm]{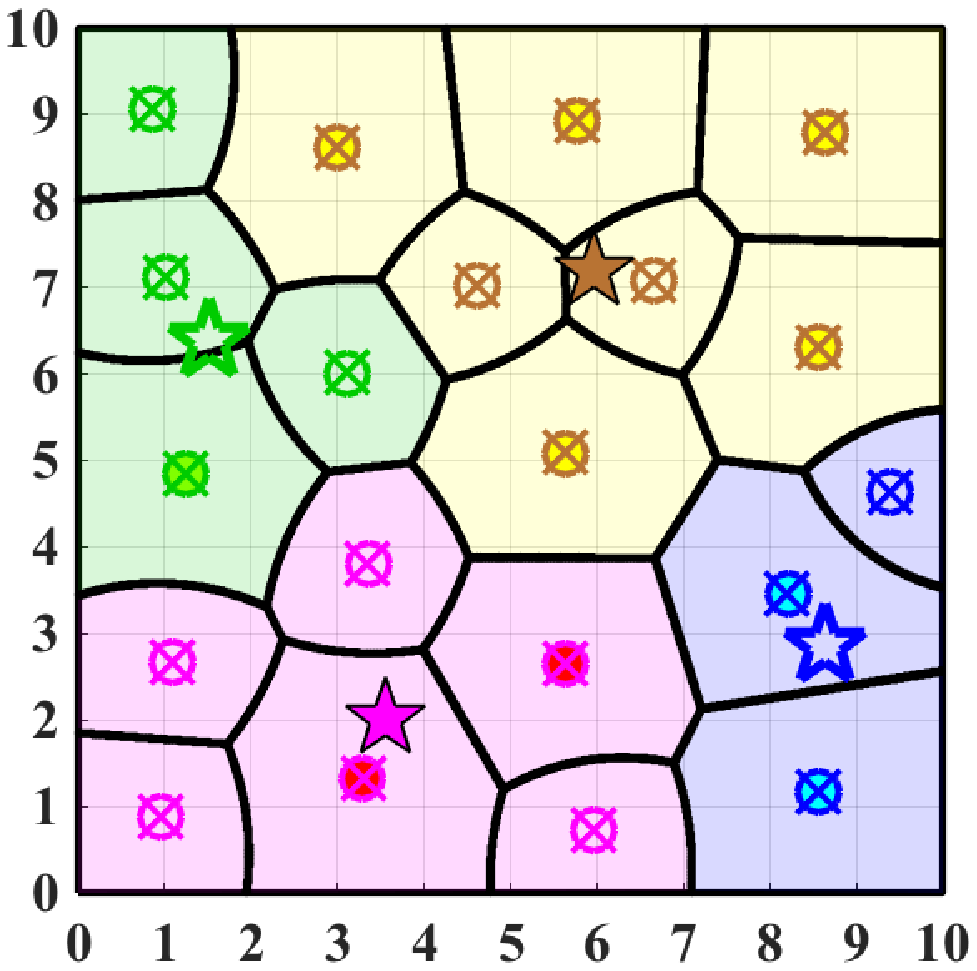}
\label{ACDeploymentInWSN2}}
\hfil
\subfloat[]{\includegraphics[width=43mm]{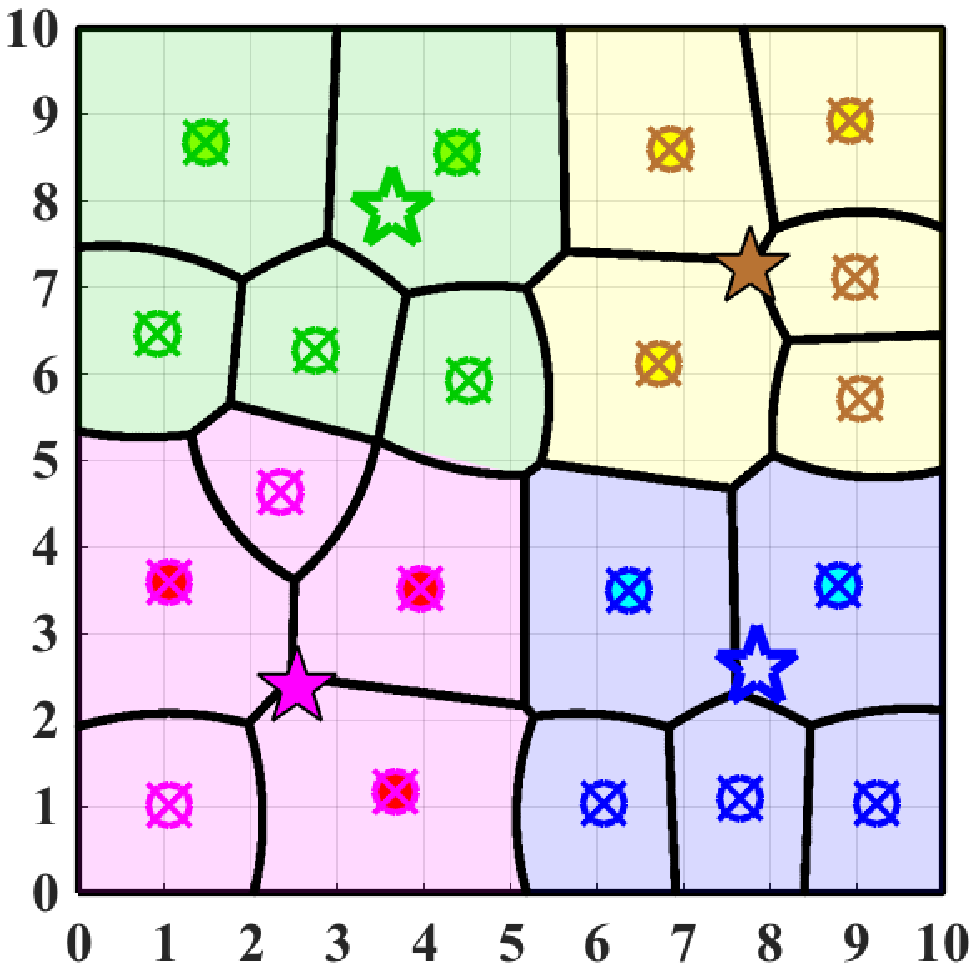}
\label{DCDeploymentInWSN2}}
\hfil
\subfloat[]{\includegraphics[width=43mm]{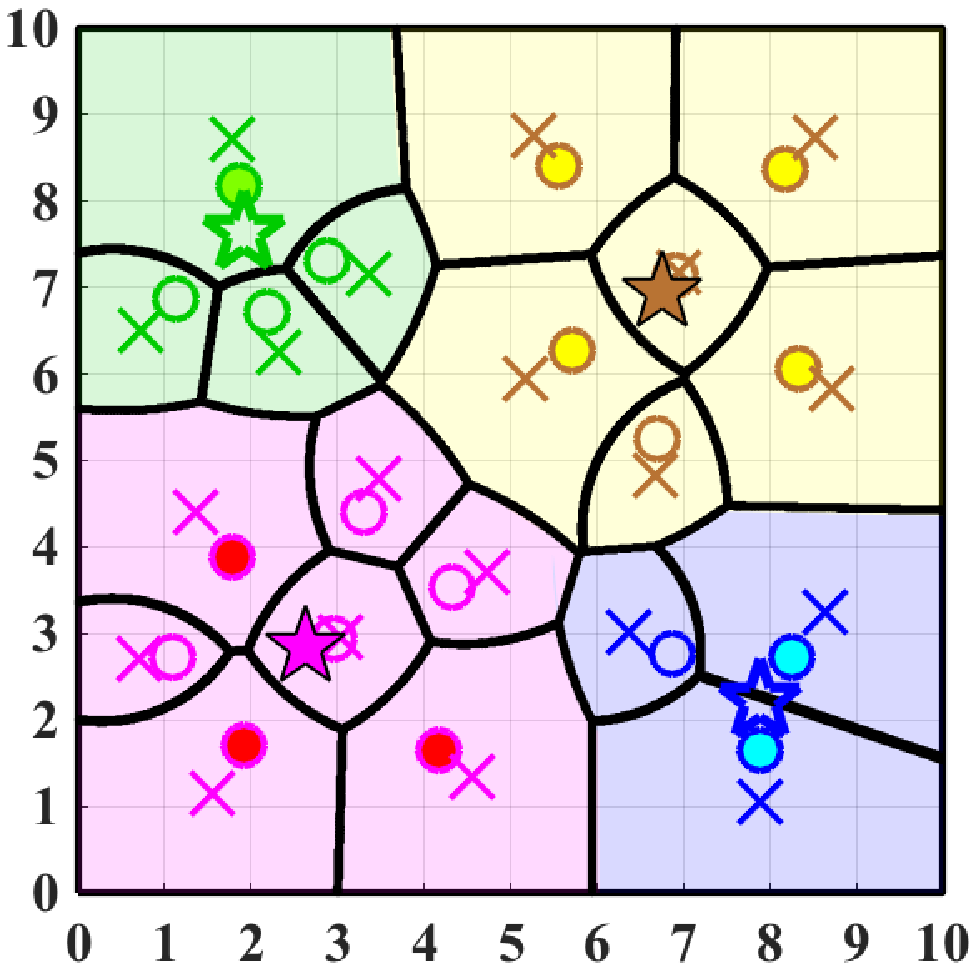}
\label{HELDeploymentInWSN2}}
\captionsetup{justification=justified}
\caption{\small{AP and FC deployments of different algorithms with $\beta=0.25$ in WSN2. (a) MER. (b) AC (c) DC. (d) HTTL.}}
\label{DeploymentInWSN2}
\end{figure}

Figs. \ref{MERDeploymentInWSN1}, \ref{ACDeploymentInWSN1}, \ref{DCDeploymentInWSN1}, and \ref{HELDeploymentInWSN1} show final deployments of the four algorithms (MER, AC, DC, and HTTL) in WSN1. The multi-hop paths are denoted by blue dotted lines. As expected from Proposition \ref{necessary}, every AP is placed on the line between the connected FC and geometric center of its cell by running HTTL Algorithm. In addition, the HTTL Algorithm deploys weak APs close to the FC while strong APs are placed on outer regions. Figs. \ref{MERDeploymentInWSN2}, \ref{ACDeploymentInWSN2}, \ref{DCDeploymentInWSN2}, and \ref{HELDeploymentInWSN2} illustrate the final deployments of MER, AC, DC, and HTTL, in WSN2, respectively. Intuitively, strong FCs provide service to more APs compared to weak FCs in both AC and HTTL Algorithms. Moreover, by HTTL Algorithm, strong APs cover larger target regions compared to weak APs in Fig. \ref{HELDeploymentInWSN2}.

\begin{figure}[!htb]
\setlength\abovecaptionskip{0pt}
\setlength\belowcaptionskip{0pt}
\centering
\subfloat[]{\includegraphics[width=43mm]{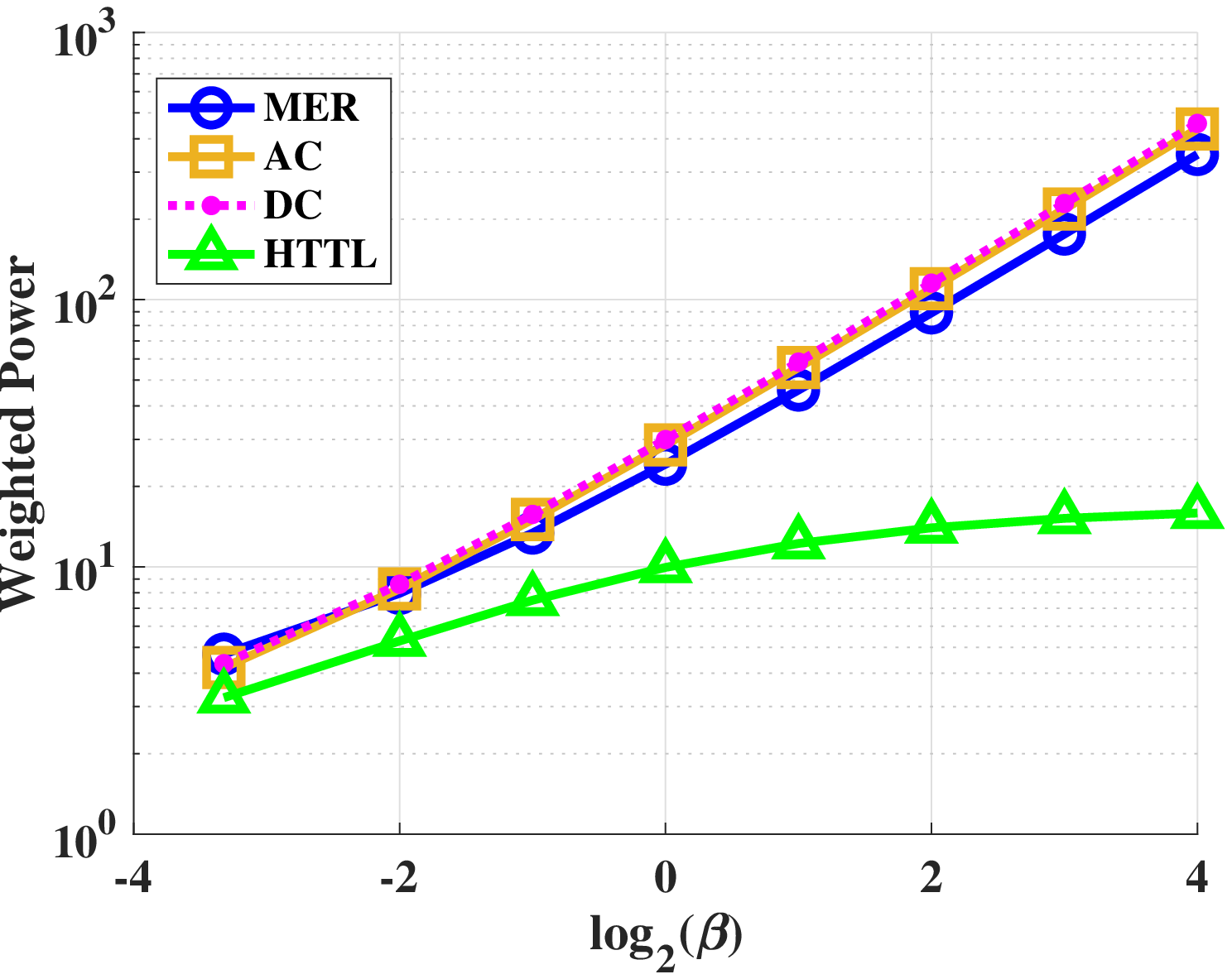}
\label{DistortionWSN1}}
\hfil
\subfloat[]{\includegraphics[width=43mm]{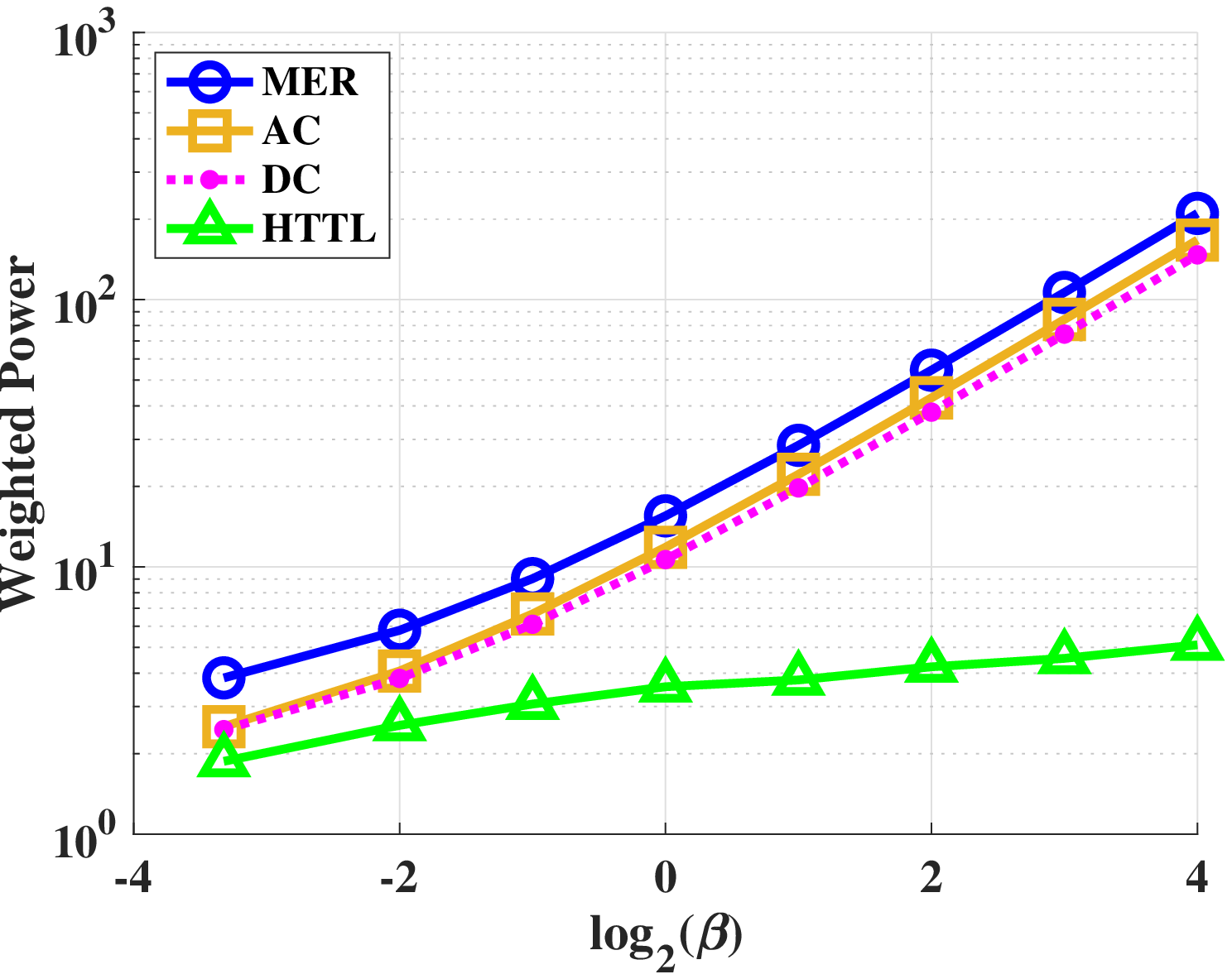}
\label{DistortionWSN2}}
\captionsetup{justification=justified}
\caption{\small{The weighted power comparison of different Algorithms. (a) WSN1. (b) WSN2.}}
\label{DistortionComparsion}
\end{figure}

Figs. \ref{DistortionWSN1} and \ref{DistortionWSN2} show the weighted power comparison of different algorithms in WSN1 and WSN2. Obviously, our proposed algorithm, HTTL, outperforms the other three algorithms in both WSN1 and WSN2. In particular, the energy consumption gap between HTTL and other three algorithms increases as the FC energy consumption becomes more important ($\beta$ increases).

\section{Conclusion}\label{sec:conclusion}

A heterogeneous two-tier network which collects data from a large-scale wireless sensor to heterogeneous fusion centers through heterogeneous access points is discussed. We studied the minimum power that ensures reliable communication on such two-tier networks and modeled it as a quantization problem. Different from the homogeneous two-tier networks, a novel Voronoi Diagram is proposed to provide the best cell partition for the heterogeneous network. The necessary conditions of optimal node deployment imply that every access point should be placed between its connected fusion center and the geometric center of its cell partition. By defining an appropriate distortion measure, we proposed a heterogeneous two-tier Lloyd Algorithm (HTTL) to minimize the distortion. Simulation results show that HTTL algorithm greatly saves the weighted power or energy in a heterogeneous two-tier network.

\section*{Appendix A}

For $U=\left(S_1,S_2,...,S_N \right)$, The left hand side of (\ref{eq7}) can be written as:
\begin{align}\label{appendixA_eq1}
D(P,Q,U,T) &= \sum_{n=1}^N\int_{S_n}(a_n||p_n-w||^2 \nonumber \\&+ \beta b_{n,T(n)}||p_n - q_{T(n)}||^2 )f(w)dw \nonumber \\ 
           &\geq \sum_{n=1}^N\int_{S_n}\min_j (a_j||p_j-w||^2 \nonumber \\&+ \beta b_{j,T(j)}||p_j - q_{T(j)}||^2 )f(w)dw \nonumber \\
           & = \int_{\Omega}\min_j (a_j||p_j-w||^2 \nonumber \\&+ \beta b_{j,T(j)}||p_j - q_{T(j)}||^2 )f(w)dw \nonumber \\
           &= \sum_{n=1}^N\int_{V_n}\min_j (a_j||p_j-w||^2 \nonumber \\&+ \beta b_{j,T(j)}||p_j - q_{T(j)}||^2 )f(w)dw \nonumber \\
           &= \sum_{n=1}^N\int_{V_n}(a_n||p_n-w||^2 \nonumber \\&+ \beta b_{n,T(n)}||p_n - q_{T(n)}||^2 )f(w)dw \nonumber \\
           &= D(P,Q, \mathbf{V}, T)
\end{align}    
Hence, the generalized Voronoi diagram is the optimal partition for any given deployment $(P,Q,T)$.$\hfill\blacksquare$

\section*{Appendix B}

Given $N$ APs and $M$ FCs $(M<N)$, first we demonstrate that there exists an optimal node deployment such as $\left(\widehat{P},\widehat{Q},\widehat{\mathbf{R}},\widehat{T} \right)$ in which each FC has at most one connected AP at the same location, i.e., for each $m\in \mathcal{I_B}$, the cardinality of the set $\{n|\widehat{T}(n)=m, \widehat{p}_n=\widehat{q}_m\}$ is less than or equal to 1. For this purpose, we consider an optimal node deployment $\left(P^*,Q^*,\mathbf{R}^*,T^* \right)$ and assume that there exist at least two distinct indices $n_1,n_2\in \mathcal{I_A}$ and an index $m\in \mathcal{I_B}$ such that $T^*(n_1)=T^*(n_2)=m$, and $p^*_{n_1}=p^*_{n_2}=q^*_m$. Without loss of generality, we can assume that $a_{n_1}\leq a_{n_2}$. We have:
\begin{align}
    D_{n_1}& =\int_{R^*_{n_1}}(a_{n_1}||p^*_{n_1}-w||^2 \nonumber \\& + \beta b_{n_1,m}||p^*_{n_1} - q^*_m||^2  )f(w)dw \nonumber \\
    &= \int_{R^*_{n_1}}\left(a_{n_1}||p^*_{n_1}-w||^2\right)f(w)dw
\end{align}
\begin{align}
    D_{n_2}& =\int_{R^*_{n_2}}(a_{n_2}||p^*_{n_2}-w||^2 \nonumber \\& + \beta b_{n_2,m}||p^*_{n_2} - q^*_m||^2  )f(w)dw \nonumber \\
    &= \int_{R^*_{n_2}}\left(a_{n_2}||p^*_{n_2}-w||^2\right)f(w)dw
\end{align}
Hence, we have:
\begin{align}\label{eq15}
    D_{n_1} + D_{n_2} &=  \int_{R^*_{n_1}}\left(a_{n_1}||p^*_{n_1}-w||^2\right)f(w)dw \nonumber\\& + \int_{R^*_{n_2}}\left(a_{n_2}||p^*_{n_2}-w||^2\right)f(w)dw \nonumber\\
& \geq \int_{R^*_{n_1}}\left(a_{n_1}||p^*_{n_1}-w||^2\right)f(w)dw \nonumber\\& + \int_{R^*_{n_2}}\left(a_{n_1}||p^*_{n_1}-w||^2\right)f(w)dw \nonumber\\
& = \int_{R^*_{n_1}\bigcup R^*_{n_2}}\left(a_{n_1}||p^*_{n_1}-w||^2\right)f(w)dw 
\end{align}
Eq. (\ref{eq15}) implies that if we update the cell partition for AP $n_1$ to be $R^*_{n_1}\bigcup R^*_{n_2}$, and place the AP $n_2$ to an arbitrary location different from $q^*_m$ with a corresponding zero volume cell partition, the resulting distortion will not increase, and the obtained node deployment is also optimal. Note that in this newly obtained optimal distortion, AP $n_2$ is not in the same location as FC $m$ anymore. This procedure is continued until we reach an optimal deployment in which each FC has at most one connected AP upon it. Let us denote this optimal node deployment via $\left(\widehat{P}, \widehat{Q},\widehat{\mathbf{R}}, \widehat{T}  \right)$.

Since $M<N$ and each FC has at most one AP upon it, there exists an index $k\in \mathcal{I_A}$ such that $\widehat{p}_k\neq  \widehat{q}_{\widehat{T}(k)}$. In order to show that the optimal two-tier distortion with $N$ APs and $M+1$ FCs is less than that of $N$ APs and $M$ FCs, it is sufficient to construct a node deployment with $N$ APs and $M+1$ FCs such as $\left(P'',Q'',\mathbf{R}'',T'' \right)$ that achieves lower distortion than $D\left(\widehat{P}, \widehat{Q},\widehat{\mathbf{R}}, \widehat{T}  \right)$. For each $n\in \mathcal{I_A}$, let $\widehat{v}_n$ denote the volume of the region $\widehat{R}_n$, i.e., $\widehat{v}_n=\int_{\widehat{R}_n}f(w)dw$. We consider two different cases: (i) If $\widehat{v}_k > 0$, then we set $P''=\widehat{P}$, $Q''=\left(\widehat{q}_1,\widehat{q}_2,...,\widehat{q}_M , q''_{M+1}=\widehat{p}_k  \right)$, $\mathbf{R}''=\widehat{\mathbf{R}}$ and $T''(n)=\widehat{T}(n)$ for $n\neq k$ and $T''(k) = M+1$. Note that
\begin{align}
    &{ }\int_{\widehat{R}_k}\left(a_k||\widehat{p}_k - w||^2 + \beta b_{k,\widehat{T}(k)} ||\widehat{p}_k - \widehat{q}_{\widehat{T}(k)}||^2   \right)f(w)dw \nonumber \\
    &> \int_{\widehat{R}_k} \left(a_k||\widehat{p}_k - w||^2 \right)f(w)dw \\
    & = \int_{\widehat{R}_k} \left(a_k||\widehat{p}_k - w||^2 + \beta b_{k,M+1}||\widehat{p}_k - q''_{M+1}||^2    \right)f(w)dw \nonumber
\end{align}
implies that in the new deployment $\left(P'',Q'',\mathbf{R}'',T'' \right)$, the contribution of the AP $k$ to the total distortion has decreased. Since the contribution of other APs to the distortion has not changed, we have $D\left(P'', Q'', \mathbf{R}'', T''  \right) < D\left(\widehat{P}, \widehat{Q},\widehat{\mathbf{R}}, \widehat{T}  \right)$ and the proof is complete. (ii) If $\widehat{v}_k = 0$, then AP $k$ does not contribute to the optimal distortion $D\left(\widehat{P},\widehat{Q},\widehat{\mathbf{R}},\widehat{T}  \right)$, and it can be placed anywhere within the target region $\Omega$. Since the set $\left\{\widehat{p}_1,...,\widehat{p}_N,\widehat{q}_1,...,\widehat{q}_M  \right\}$ has zero measure, clearly there exists a point $x\in \Omega$ and a threshold $\delta \in \mathbb{R}^+$ such that $\mathcal{B}\left(x,\delta \right)=\left\{w\in \Omega | \|x-w\|\leq \delta \right\}$ does not include any point from the set $\left\{\widehat{p}_1,...,\widehat{p}_N,\widehat{q}_1,...,\widehat{q}_M  \right\}$. Since $f(.)$ is positive, continuous and differentiable over $\Omega$, for each $0 < \epsilon < \delta$ the region $\mathcal{B}(x,\epsilon)=\left \{w\in \Omega \big| ||w-x||\leq \epsilon   \right\}$ has positive volume, i.e., $\int_{\mathcal{B}(x,\delta)}f(w)dw > 0$. Given $0< \epsilon < \delta$, assume that:
\begin{equation}\label{eq18.5}
    \mathcal{B}(x,\epsilon)\subset \widehat{R}_n
\end{equation}
for some $n\in \mathcal{I_A}$; therefore, the contribution of the region $\mathcal{B}(x,\epsilon)$ to the total distortion $D\left(\widehat{P},\widehat{Q},\widehat{\mathbf{R}},\widehat{T} \right)$ is equal to:
\begin{equation}\label{eq19}
    \int_{\mathcal{B}(x,\epsilon)}\left(a_n||\widehat{p}_n - w||^2 + \beta b_{n,\widehat{T}(n)}||\widehat{p}_n - \widehat{q}_{\widehat{T}(n)}||^2        \right)f(w)dw
\end{equation}
As $\epsilon \longrightarrow 0$, (\ref{eq19}) can be approximated as:
\begin{equation}\label{eq20}
  \Delta_n \times   \int_{\mathcal{B}(x,\epsilon)}f(w)dw
\end{equation}
where $\Delta_n = \left(a_n||\widehat{p}_n - x||^2 + \beta b_{n,\widehat{T}(n)}||\widehat{p}_n - \widehat{q}_{\widehat{T}(n)}||^2        \right)$. If we set $p''_k=q''_{M+1}=x$ and $R''_k = \mathcal{B}(x,\epsilon)$ and $T''(k)=M+1$, then the contribution of the region $\mathcal{B}(x,\epsilon)$ to the total distortion $D\left(P'',Q'',\mathbf{R}'',T'' \right)$ is equal to:
\begin{align}\label{eq21}
    &{}\int_{\mathcal{B}(x,\epsilon)}\left(a_k||p''_k-w||^2 + \beta b_{k,M+1}||p''_k - q''_{M+1}||^2 \right)f(w)dw \nonumber \\
    &= a_k \int_{\mathcal{B}(x,\epsilon)}\left(||x - w||^2 \right)f(w)dw
\end{align}
The below equation for the ratio of distortions in (\ref{eq20}) and (\ref{eq21})
\begin{equation}\label{eq23}
    \lim_{\epsilon \longrightarrow 0}\frac{a_k \int_{\mathcal{B}(x,\epsilon)}\left(||x - w||^2 \right)f(w)dw}{\Delta_n \times   \int_{\mathcal{B}(x,\epsilon)}f(w)dw} = 0
\end{equation}
implies that there exists an $\epsilon^*\in (0,\delta)$ such that the contribution of the region $\mathcal{B}(x,\epsilon^*)$ to the total distortion in $D\left( P'', Q'', \mathbf{R}'', T'' \right)$ will be less than that of $D\left(\widehat{P}, \widehat{Q}, \widehat{\mathbf{R}}, \widehat{T}\right)$. Hence, we set $P''=\left(p''_1,p''_2,...,p''_N \right)$ where $p''_i = \widehat{p}_i$ for $i\neq k$, and $p''_k = x$. Also, we set $Q''=\left(\widehat{q}_1,\widehat{q}_2,...,\widehat{q}_M,q''_{M+1} = x   \right)$. The partitioning $\mathbf{R}''=\left(R''_1,...,R''_N   \right)$ is defined as $R''_i=\widehat{R}_i$ for $i\neq k$ and $i\neq n$, $R''_k = \mathcal{B}(x,\epsilon^*)$ and $R''_n = \widehat{R}_n - \mathcal{B}(x,\epsilon^*)$. Finally, we set $T''(i) = \widehat{T}(i)$ for $i\neq k$ and $T''(k) = M+1$. As mentioned earlier, the two-tier distortion $D\left(P'',Q'',\mathbf{R}'',T''   \right)$ is less than $D\left (\widehat{P}, \widehat{Q}, \widehat{\mathbf{R}}, \widehat{T}    \right)$. Note that if the region $\mathcal{B}(x,\epsilon)$ is a subset of more than one region, Eqs. (\ref{eq18.5}) to (\ref{eq20}) and (\ref{eq23}) can be modified accordingly and a similar argument can be made to show that the resulting distortion will be improved in the new deployment, and the proof is complete.$\hfill\blacksquare$

\section*{Appendix C}

Assume that there exists an index $m\in \mathcal{I_B}$ in the optimal node deployment $\left(P^*, Q^*, \mathbf{R}^*, T^* \right)$ such that $\bigcup_{n:T^*(n)=m}R^*_n$ has zero volume. Consider the node deployment $\left(P',Q',\mathbf{R}',T' \right)$ where $P'=P^*$, $Q' =  \left(q^*_1,...,q^*_{m-1},q^*_{m+1},...,q^*_M\right)$, $\mathbf{R}' = \mathbf{R}^*$ and $T'(i)=T^*(i)$ for indices $i\in \mathcal{I_A}$ such that $T^*(i)\neq m$. Note that for indices $i\in \mathcal{I_A}$ such that $T^*(i)=m$, we can define $T'(i)$ arbitrarily because the corresponding regions $R'_i$ have zero volume. Since $\bigcup_{n:T^*(n)=m}R^*_n$ has zero volume, we have:
\begin{equation}\label{eq24}
    D\left (  P',Q',\mathbf{R}',T'  \right) = D\left (  P^*,Q^*,\mathbf{R}^*,T^*  \right)
\end{equation}
which is in contradiction with Lemma 1 since the optimal node deployment $\left(P^*, Q^*, \mathbf{R}^*, T^* \right)$ for $N$ APs and $M$ FCs has not improved the node deployment $\left(P',Q',\mathbf{R}',T' \right)$ for $N$ APs and $M-1$ FCs in terms of distortion. Hence the proof is complete.$\hfill\blacksquare$

\section*{Appendix D}

First, we study the shape of the Voronoi regions in (\ref{eq5}). Let $\mathcal{B}(c,r)=\{\omega|\|\omega-c\|\leq r\}$ be a disk centered at $c$ with radius $r$ in two-dimensional space. In particular, $\mathcal{B}(c,r)=\emptyset$ when $r\leq 0$.
Let $\mathcal{HS}=\{\omega|A\omega+B\leq0\}$ be a half space, where $A\in\mathbb{R}^2$ is a vector and $B\in\mathbb{R}$ is a constant. 
For $i,j \in \mathcal{I_A}$, we define 
\begin{equation}
\begin{aligned}
    V_{ij}(P,Q,T)\triangleq\{\omega|a_i||p_i-w||^2 + \beta b_{i, T(i)}||p_i-q_{T(i)}||^2 \leq \\ a_j||p_j-w||^2 + \beta b_{j, T(j)}||p_j-q_{T(j)}||^2{\color{black}\}}
\end{aligned}
\label{Vij}
\end{equation}
to be the pairwise Voronoi region of AP $i$ where only AP $i$ and $j$ are considered. Then, AP $i$'s Voronoi region can be represented as $V_i(P, Q) = \left[\bigcap_{j\neq i}V_{ij}(P,Q)\right]\bigcap\Omega$.
Let $(\omega_x,\omega_y)$, $(p_{ix}, p_{iy})$, and $(p_{jx}, p_{jy})$ be the coordinates of $\omega$, $p_i$ and $p_j$, respectively. 
Expanding the inequality in (\ref{Vij}) results in 
\begin{equation}
\begin{aligned}
    &(a_i-a_j)(\omega_x^2+\omega_y^2)-2(a_ip_{ix}-a_jp_{jx})\omega_x\\
    &-2(a_ip_{iy}-a_jp_{jy})\omega_y
    +a_i\|p_i\|^2-a_j\|p_j\|^2\\
    &+\beta b_{i,T(i)}\|p_i\!-\!q_{T(i)}\|^2
    -\beta b_{j,T(j)}\|p_j\!-\!q_{T(j)}\|^2{\color{black}\leq}0
\end{aligned}
\end{equation}
When $a_i=a_j$, the pairwise Voronoi region is a half space, i.e., $V_{ij} = \{A_{ij}\omega + B_{ij} \leq 0\}$, where $A_{ij}=a_jp_j-a_ip_i$ and $        B_{ij} = \frac{\left(a_i||p_i||^2-a_j||p_j||^2+\beta b_{i, T(i)}||p_i-q_{T(i)}||^2-\beta b_{j, T(j)}||p_j-q_{T(j)}||^2\right)}{2}$.
When $a_i>a_j$, $V_{ij}$ is represented as:
\begin{equation}
    (\omega - c_{ij})^2 \leq L_{ij}.
\end{equation}
When $a_i<a_j$, $V_{ij}$ is represented as:
\begin{equation}
    (\omega - c_{ij})^2 \geq L_{ij},
\end{equation}
where 
\begin{align}
     c_{ij}&=\left(\frac{a_ip_{ix}-a_jp_{jx}}{a_i-a_j}, \frac{a_ip_{iy}-a_jp_{jy}}{a_i-a_j}\right)=\frac{a_ip_i-a_jp_j}{a_i-a_j} \\
     L_{ij}&=\frac{a_ia_j\|p_i-p_j\|^2}{\left(a_i-a_j\right)^2} \nonumber \\&-\beta \times \frac{ b_{i, T(i)}\|p_i\!-\!q_{T(i)}\|^2- b_{j, T(j)}\|p_j\!-\!q_{T(j)}\|^2}{(a_i-a_j)}
 \end{align} 

For $L_{ij}\geq 0$, we define the radius $r_{ij}$ as:
\begin{equation}\label{r_i_j}
            r_{ij} = \begin{cases}
    \sqrt{L_{ij}}, &\mbox{$L_{ij} \geq 0$}\\
    0, &\mbox{$L_{ij} < 0$}
    \end{cases}
\end{equation}
Therefore, the pairwise Voronoi region $V_{ij}$ is derived:
\begin{align}\label{pairwise_V_i_j}
    V_{ij}=\Omega\cap\begin{cases}
         \mathcal{HS}(A_{ij},B_{ij})&, a_i=a_j  \\
         \mathcal{B}(c_{ij},r_{ij}) &, a_i>a_j , L_{ij}\geq 0\\
                  \varnothing &, a_i>a_j , L_{ij}< 0\\
         \mathcal{B}^c(c_{ij},r_{ij}) &, a_i<a_j , L_{ij} \geq 0 \\
         \mathbb{R}^2 &, a_i<a_j , L_{ij}< 0
        \end{cases},
  \end{align}
where $\mathcal{B}^c(c_{ij},r_{ij})$ is the complementary of  $\mathcal{B}(c_{ij},r_{ij})$. Note that for two distinct indices such as $i,j\in \mathcal{I_A}$, if $a_i>a_j$ and $L_{ij}<0$, then two regions $\Omega\cap\mathcal{B}(c_{ij},r_{ij})$ and $\varnothing$ differ only in one point, i.e., $c_{ij}$. Similarly, for $a_i<a_j$ and $L_{ij}<0$, two regions $\Omega\cap\mathcal{B}^c(c_{ij},r_{ij})$ and $\Omega$ differ only in one point $c_{ij}$. Hence, if we define:
\begin{align}\label{voronoi_v_n}
    \overline{V}_k = &\left [ \bigcap_{i:a_k>a_i} \mathcal{B}(c_{ki},r_{ki})  \right] \bigcap \nonumber \\ 
    &\left[\bigcap_{i:a_k=a_i} \mathcal{HS}(A_{ki},B_{ki}) \right] \bigcap \nonumber \\
    &\left[ \bigcap_{i:a_k<a_i} \mathcal{B}^c(c_{ki},r_{ki})  \right] \bigcap \Omega
\end{align}
then two regions $\overline{V}_k$ and $V_k$ differ only in finite number of points. As a result, integrals over both $\overline{V}_k$ and $V_k$ have the same value since the density function $f$ is continuous and differentiable, and removing finite number of points from the integral region does not change the integral value. Note that if $V_k$ is empty, the Proposition 1 in \cite{JG2} holds since the integral over an empty region is zero. If $V_k$ is not empty, the same arguments as in Appendix A of \cite{JG2} can be replicated since $\overline{V}_k$ in (\ref{voronoi_v_n}) is similar to Eq. (31) in \cite{JG2}.

Using parallel axis theorem, the two-tier distortion can be written as:
\begin{align}\label{distortion_parallel_axis_theorem}
    D\left(P,Q,\mathbf{V},T \right) &= \sum_{n=1}^{N}\int_{V_n}\bigg(a_n||p_n - w||^2 \nonumber \\& + \beta b_{n,T(n)}||p_n - q_{T(n)}||^2  \bigg)f(w)dw \nonumber \\
    &= \sum_{n=1}^{N}\bigg( \int_{V_n}a_n||c_n-w||^2f(w)dw \nonumber \\
    &+ a_n||p_n-c_n||^2v_n \nonumber \\
    &+\beta b_{n,T(n)}||p_n-q_{T(n)}||^2v_n \bigg)
\end{align}
Using Proposition 1 in \cite{JG2}, since the optimal deployment $\left(P^*,Q^*  \right)$ satisfies zero gradient, we take the partial derivatives of Eq. (\ref{distortion_parallel_axis_theorem}) as follows:
\begin{align}\label{partial_derivative}
    \frac{\partial D}{\partial p^*_n} & = 2\left[ a_n(p^*_n - c^*_n) + \beta b_{n,T^*(n)}(p^*_n - q^*_{T^*(n)}) \right]v^*_n  =0 \nonumber \\
    \frac{\partial D}{\partial q^*_m}& = 2\sum_{n:T^*(n)=m} \beta b_{n,m}(q^*_m - p^*_n)v^*_n = 0
\end{align}
By solving Eq. (\ref{partial_derivative}), we have the following necessary conditions:
\begin{align}\label{necesasry_condition}
    p_n^* &= \frac{a_nc^*_n + \beta b_{n,T^*(n)}q^*_{T^*(n)}}{a_n + \beta b_{n,T^*(n)}} \\
    q^*_m &= \frac{\sum_{n:T^*(n)=m}b_{n,m}p^*_nv^*_n}{\sum_{n:T^*(n)=m}b_{n,m}v^*_n}
\end{align}
and the proof is complete.$\hfill\blacksquare$

\section*{Appendix E}

In what follows, we demonstrate that none of the four steps in the HTTL algorithm will increase the two-tier distortion. Given $P$, $Q$ and $\mathbf{R}$, updating the index map $T$ according to (\ref{eq8}) minimizes the total distortion, i.e., the two-tier distortion will not increase by the first step. Moreover, given $P$, $Q$ and $T$, Proposition \ref{allactive} indicates that updating $\mathbf{R}$ according to (\ref{eq5}) and (\ref{eq6}) gives the best partitioning; thus, the second step of the HTTL algorithm will not increase the distortion. Below equality follows from straightforward algebraic calculations and we omit the proof here:
\begin{align}\label{appendix_F_1}
    \sum_{n:T(n)=m}b_{n,m}v_n||p_n - q_m||^2 = &\sum_{n:T(n)=m}b_{n,m}v_n(||p_n - q'_m||^2 \nonumber \\&+ ||q_m - q'_m||^2 )
\end{align}
for $q'_m=\frac{\sum_{n:T(n)=m}b_{n,m}p_nv_n}{\sum_{n:T(n)=m}b_{n,m}v_n}$. The contribution of FC $m$ to the total distortion can then be rewritten as:
\begin{align}\label{appendix_F_2}
    &\sum_{n:T(n)=m}\int_{R_n}\left(a_n||p_n-w||^2 + \beta b_{n,m}||p_n-q_m||^2\right)f(w)dw \nonumber \\&=
    \sum_{n:T(n)=m}\int_{R_n}a_n||p_n-w||^2f(w)dw \nonumber \\&+ \beta \left(\sum_{n:T(n)=m}b_{n,m}v_n  \right)||q_m - q'_m||^2 \nonumber \\&+
    \beta \left(\sum_{n:T(n)=m}b_{n,m}v_n||p_n - q'_m||^2 \right)
\end{align}
Now, given $P$, $\mathbf{R}$ and $T$, the first and third terms in right hand side of (\ref{appendix_F_2}) are constant and moving $q_m$ toward $q'_m$ will not increase the distortion in (\ref{appendix_F_2}). Therefore, the third step of the HTTL algorithm will not increase the total two-tier distortion as well.

The following equation can be easily verified using straightforward algebraic computations and we omit the proof here:
\begin{align}\label{appendix_F_3}
   & a_n||p_n-w||^2 + \beta b_{n,m}||p_n - q_m||^2 \nonumber \\& = (a_n + \beta b_{n,m})\left|\left|p_n - \frac{(a_nw + \beta b_{n,m}q_m)}{a_n + \beta b_{n,m}}\right|\right|^2 \nonumber \\& + \frac{\beta a_n b_{n,m}}{a_n + \beta b_{n,m}}||w-q_m||^2
\end{align}
For each index $n\in \mathcal{I_A}$ and the corresponding index $m=T(n)$, we can rewrite the contribution of AP $n$ to the total distortion as:
\begin{align}\label{appendix_F_4}
    & \int_{R_n}\left(a_n||p_n-w||^2 + \beta b_{n,m}||p_n - q_m||^2  \right)f(w)dw \nonumber \\
    &=\int_{R_n}\bigg[\left(a_n + \beta b_{n,m}\right)\left|\left|p_n - \frac{\left(a_nw + \beta b_{n,m}q_m\right)}{a_n + \beta b_{n,m}}\right|\right|^2  \nonumber \\ \nonumber  &+  \frac{\beta a_n b_{n,m}}{a_n + \beta b_{n,m}}||w-q_m||^2 \bigg]f(w)dw \nonumber\\ &=
    \int_{R_n}\bigg[ \frac{a_n^2}{a_n+\beta b_{n,m}}\bigg|\bigg| \frac{\left( a_n + \beta b_{n,m} \right)p_n - \beta b_{n,m}q_m}{a_n}  - w \bigg|\bigg|^2 \nonumber \\&+ \frac{\beta a_nb_{n,m}}{a_n + \beta b_{n,m}}||w - q_m||^2 \bigg]f(w)dw \nonumber \\& =
     \int_{R_n}\bigg[ \frac{a_n^2}{a_n+\beta b_{n,m}}\bigg(\bigg|\bigg| \frac{\left( a_n + \beta b_{n,m} \right)p_n - \beta b_{n,m}q_m}{a_n}  - c_n \bigg|\bigg|^2 \nonumber \\&+ ||c_n - w||^2\bigg) +  \frac{\beta a_nb_{n,m}}{a_n + \beta b_{n,m}}||w - q_m||^2 \bigg]f(w)dw \nonumber \\& =
      \int_{R_n}\bigg[\frac{a_n^2}{a_n+\beta b_{n,m}}||c_n - w||^2 \nonumber \\&+ (a_n+\beta b_{n,m})\bigg|\bigg|p_n - \frac{a_nc_n + \beta b_{n,m}q_m}{a_n + \beta b_{n,m}}   \bigg|\bigg|^2 \nonumber \\&+ \frac{\beta a_nb_{n,m}}{a_n + \beta b_{n,m}}||w-q_m||^2 \bigg]f(w)dw \nonumber \\&= 
      \frac{a_n^2}{a_n+\beta b_{n,m}}\int_{R_n}||c_n - w||^2f(w)dw \nonumber \\&+ (a_n+\beta b_{n,m})||p_n - p'_n   ||^2 v_n \nonumber \\&+ \frac{\beta a_nb_{n,m}}{a_n + \beta b_{n,m}}\int_{R_n}||w-q_m||^2 f(w)dw
\end{align}
where $p'_n = \frac{a_nc_n + \beta b_{n,m}q_m}{a_n + \beta b_{n,m}}$. Note that the first equality in (\ref{appendix_F_4}) comes from (\ref{appendix_F_3}), and the third equality follows from the parallel axis theorem. Now, given $Q$, $\mathbf{R}$ and $T$, the first and third terms in (\ref{appendix_F_4}) are constant and moving $p_n$ toward $p'_n$ will not increase the second term in (\ref{appendix_F_4}). Hence, the fourth step of the HTTL algorithm will not increase the total distortion either. So, the HTTL algorithm generates a sequence of positive non-increasing distortion values and thus, it converges. Note that if distortion remains the same after an iteration of the algorithm, it means that non of the four steps have decreased distortion and the algorithm has already reached an optimal deployment.$\hfill\blacksquare$

\section*{Acknowledgment}
This work was supported in part by the NSF Award CCF-1815339.


\end{document}